\documentclass[12pt]{article}
\usepackage[dvips]{graphicx}
\usepackage{amsmath,amssymb}

  \newcommand{\be}{\begin{equation}}
\newcommand{\ee}{\end{equation}}

\newcommand{\vek}{\mbox{\boldmath${\rm k}$}}

\newcommand{\ver}{\mbox{\boldmath${\rm r}$}}
\newcommand{\vesig}{\mbox{\boldmath${\rm \sigma}$}}
\newcommand{\vep}{\mbox{\boldmath${\rm p}$}}
\newcommand{\veq}{\mbox{\boldmath${\rm q}$}}

\newcommand{\veA}{\mbox{\boldmath${\rm A}$}}

\newcommand{\vegamma}{\mbox{\boldmath${\rm \gamma}$}}

\newcommand{\lan}{\langle}
\newcommand{\ran}{\rangle}

\DeclareMathOperator{\grad}{grad}

\begin{document}

\title{Toward the Understanding of Quark Matter Formation}

\author{B.Kerbikov and E.V.Luschevskaya\\
 State Research
Center\\Institute of Theoretical and Experimental Physics, \\
Moscow, Russia}

 \date{}

\maketitle

\begin{abstract}
We present the evidence that crossover, fluctuation phenomena and
possible Anderson transition are the precursors  of quark matter formation.
\end{abstract}
\section{Introduction}

During the last decade, the investigation of quark matter at
finite temperature and density has become one of the  QCD focal
points. It is expected that at densities which are 3-5 times
larger than the  normal nuclear density baryons are crushed into
quarks. It is also expected that if the temperature of such quark
matter is low enough (below few tens of MeV) the system is
unstable with respect to the formation of quark-quark Cooper-pair
condensate \cite{1}-\cite{3}. This phenomenon is called color
superconductivity since diquarks belong to the $\bar 3$ color
channel. At present there is a fair understanding of color
superconductivity physics in the regime of ultra-high density when
$\alpha_s$ is small \cite{4}. Most interesting is, however, the
region of moderate densities (3-5 times larger than the normal
nuclear density). This region is important for physics of neutron
stars and may possibly be investigated in the laboratory, in
particular, in future experiments at the GSI heavy-ion
machine\footnote{In heavy-ion collisions high-density state is
naturally formed with high temperature thus preventing diquark
condensation.}. In the moderate density/strong coupling regime the
theory faces the well-known difficulties of the nonperturbative
QCD. Lattice QCD calculations encounter serious obstacle at
nonzero density since in this case the determinant of the Dirac
operator is complex resulting in non-positive measure of the
corresponding path integral. Still, several attempts to perform
lattice calculations at nonzero density have been performed -- see
the review paper \cite{5}. The region of moderate densities has
been extensively studied within the framework of the
Nambu-Jona-Lasinio (NJL) type models (see \cite{6,7} and
references therein), using the instanton gas model \cite{8}, or
chiral perturbation theory \cite{9}.

The main conclusion reached within the NJL-type models is that
transition from the nuclear matter (NM) phase to the quark matter
(QM) phase occurs very early, namely when the quark density
reaches the  value only three times larger than the density of
quarks in normal nuclear matter. This  corresponds to the value
$\mu \simeq 400 $~MeV of the quark chemical potential. In the
NJL-type models the $(NM)\to (QM)$ transition has two main
signatures, namely:

(i) The gap equation for the diquark channel acquires a nontrivial
solution.

(ii) The quark constituent mass tends to zero\footnote{In the
density region under consideration only $u$- and $d$-quarks
participate in possible pairing} and the chiral symmetry is
restored. At $T=0$ the transition is believed to be of the first
order.

Both conclusions should be taken with reservations due to
oversimplifications  inherent for the models and certain
arbitrariness in the interpretation of the results.

The principal  deficiency of the existing approaches to the
$(NM)\to (QM)$ transition is the lack of understanding what
happens to the gluon sector when the density increases. From
thermodynamic arguments it follows that possible color diquark
condensate and gluon condensate have the same energy scale which
results in competition between them somewhat similar to Meissner
effect \cite{10}. It also has been demonstrated within the
Ginzburg-Landau approach (see below) that fluctuations of the
gluon field are important and may significantly influence the
character of the phase transition and the critical temperature
\cite{1,11,12}. From the fact that the would-be diquark condensate
belongs to color anti-triplet it follows that 5 of 8 gluons become
massive \cite{3}. However, first principle derivation of the QCD
string and gluon condensate evolution with the increase of the
density is not currently available. An investigation aimed at the
resolution of these problems has been attempted very recently
\cite{13}.

In the present paper we specify the set of the key parameters
which characterize the $(NM)\to (QM)$ transition and estimate the
values of these parameters. In this way we obtain a
model-independent though schematic picture which exhibits several
nontrivial phenomena. The onset of the quark phase and its further
evolution to higher densities may be viewed  as a crossover from
the strong coupling regime of composite nonoverlapping bosons
(diquarks) to the weak coupling regime of macroscopic overlapping
Cooper pair condensate\footnote{This is a particular case of the
Bose-Einstein condensate to Bardeen-Cooper-Schrieffer (BEC-BCS)
crossover -- see below.}. Another feature of the transition region
is the  drastic increase of fluctuations.  The $(NM)\to (QM)$
evolution possesses also some features of the Anderson transition.
All these three properties are interrelated.

\section{BEC-BCS Crossover and the \\
Ginzburg-Levanyuk Number}

As already mentioned  in the Introduction, the gap equation for
the diquark channel derived within the framework of the NJL-type
models acquires a nontrivial solution (a gap) at  $\mu\simeq 0.4$
GeV \cite{6,7}. To describe the crossover we will need the
corresponding value of the quark number density. Relation between
$\mu$ and $n$ is given by the well-known equation \be
n=-\frac{\partial \Omega}{\partial \mu }, \label{1} \ee where
$\Omega (T, \mu, \Delta)$ is the thermodynamic potential, $\Delta
$ - the gap parameter. For the NJL-type of models $\Omega$ is
easily calculated \cite{6,7} (see below). According to \cite{6}
for $N_f=2$ and $T=0$ transition to $\Delta\neq 0$ phase occurs at
$\mu=0.292$ GeV \footnote{Strange quark starts to participate in
pairing an much  higher densities when $\mu\gg m_s \simeq 150$
MeV.}. Then  Eq. (\ref{1}) yields $n^{1/3} =0.18 $ GeV \cite{6}.
This number was obtained in the chiral limit , i.e. under the
condition that the quark constituent mass goes to zero when
$\Delta \neq 0$. In our view the conclusion that $m=0$ when
$\Delta \neq 0  $ may be a specific feature of the NJL model (or
some versions of this model). The behavior of the quark
constituent mass throughout the $(NM)\to (QM)$ transition is still
a moot point.

The above result for $n$ may be with a good accuracy reproduced
using  the equation  for free degenerate quarks \be n=N_c N_f
\frac{k^3_F}{3\pi^2}=\frac{2}{\pi^2} k^3_F \simeq
\frac{2}{\pi^2}\mu^3,\label{2}\ee where $\mu=\left(
k^2_F+m^2\right)^{1/2}$, and $\mu=k_F$ in the chiral limit. For
$\mu=0.292$ GeV Eq.(\ref{2}) yields $n^{1/3} =0.17$ GeV in a good
agreement with Eq.(\ref{1}). For $\mu=0.4$ GeV Eq.(\ref{2}) gives
$n^{1/3}=0.23$ GeV. Thus we conclude that $n^{1/3}\simeq 0.2$ GeV
$\simeq 1$ fm$^{-1}$ in the  transition region.

We now turn our  attention to the physics behind the fact that the
gap equation derived within the framework of the NJL model and in
the mean-field approximation acquires a nontrivial solution
starting from $\mu\simeq 0.4$ GeV. It took quite some time before
it was realized \cite{14,15} that a nonzero value of the gap does
not mean the onset of the  color superconductivity (the BCS
regime). It is only a signal of the presence of fermion pairs.
Depending on the dynamics of the system, on the fermion density,
and on the  temperature, such pairs may be either stable, or
fluctuating in time, may form a BCS condensate, or a dilute Bose
gas, or undergo a Bose-Einstein (BE)  condensation. The fact that
there is a continuous transition (crossover) from the strong
coupling/low density regime of independent bound state formation
to the  weak coupling/high density cooperative Cooper pairing is
well known\cite{16}. In contrast to macroscopic Cooper pairs, the
compact molecular -- like states which are formed in strong
coupling/low density regime are called Schafroth pairs
\cite{17,18}.

The dimensionless crossover parameter is $n^{1/3}\xi$ , where
$\xi$ is the characteristic length of pair correlation when the
system is in the BCS regime and the root of the mean square radius
of the bound state when the system is in the strong coupling
regime. Some arbitrariness occurs in the definition of the
crossover parameter. For example, in \cite{19} it is defined as
$k_F\xi \simeq 1.7 n^{1/3}\xi$ (see (\ref{2})). Another definition
of the crossover parameter is $x_0 =\mu/\Delta $  \cite{20}.

In the BCS theory $\xi$ is given by $\xi=v_F/\pi\Delta$, where
$v_F$ is the velocity at the Fermi surface. For a typical metal
superconductor $v_F \simeq c/137, ~~ \Delta\simeq 5K$, so  that
$\xi\simeq 10^{-4}$ cm. The density of electrons is $n\simeq
10^{22}$ cm$^{-3}$. Therefore in the BCS regime $n^{1/3} \xi
\gtrsim 10^3$. In coordinate space the wave function of the Cooper
pair is proportional to  $(\sin k_Fr/k_Fr)\exp (-r/\xi)$ and hence
it has $\sim 10^3$ nodes.

The crossover from the BCS to the strong coupling regime occurs at
\cite{16}, \cite{19}-\cite{22} \be n^{1/3}\xi \sim 1.\label{3}\ee
The width of the crossover region with respect to the above
parameter is several units and is model-dependent
\cite{19}-\cite{22}. It was first pointed out in \cite{14,15} that
at $\mu\sim (0.3-0.5)$ GeV the quark system is in the crossover
regime and not in the BCS regime as it was inferred from the fact
that at such values  of $\mu$ the gap equation acquires a
nontrivial solution.

Let us estimate the value of the crossover parameter at the onset
of the phase with $\Delta \neq 0$, i.e.,  at $\mu\sim (0.3-0.5)$
GeV. We have seen that $n^{1/3} \simeq 1$~fm$^{-1}$ in this
region\footnote{In NJL -type models transition to $\Delta \neq 0$
phase is accompanied by a sharp increase of density towards the
value indicated above. Fluctuation corrections should smoother
this behavior.}. The value of $\xi$ in the strong coupling regime
cannot be evaluated from the first principles. One may expect that
it is of a typical hadronic scale $\xi\sim(1-2)$~fm and that it grows
with density asymptotically approaching the BCS value. Model
calculations confirm this expectations \cite{23} -\cite{25}. At
zero density the root-mean square radius of the diquark is $\simeq
1$~fm \cite{23}.  At low density the single-gluon exchange model
leads to the result $\xi \simeq n^{-1/3}$ \cite{25}, while at
$\mu\simeq 1$ GeV $\xi n^{1/3} \simeq 10 $ \cite{25}. It should be
noted that the calculation of the pair size in the crossover
region is a complicated task even for the ``simpler'' system of
ultracold fermionic atoms\cite{26}. To sum up, we conclude that
$\xi n^{1/3} \sim  1$ at the onset of the $\Delta \neq 0$ phase.
The $(NM) \to (QM)$ transition brings the system to the quark
matter state in the crossover regime, but not in  color
superconducting state.

Next we turn to the structure of the diquark wave function in the
strong coupling regime at nonzero density. The corresponding
relativistic equation has been derived in \cite{21}. The pair wave
function has the following asymptotic form \be \psi \sim
\frac{1}{r} \exp \left\{ -\left[
(\mu+m)\frac{\varepsilon_b}{2}\right]^{1/2} r
\right\},\label{4}\ee where the binding energy is given by
$\varepsilon_b/2=m-\mu$. If we introduce the nonrelativistic
chemical potential $\tilde \mu =\mu -m$, we get $\varepsilon_b =-
2\tilde \mu$ in complete agreement with the result obtained by
Nozieres and Schmitt-Rink\cite{16} (recall that $\tilde \mu<0$ in
the strong coupling limit). In the nonrelativistic case we may
also write $\mu+m =\tilde \mu +2m \simeq 2m$, and then (\ref{4})
reduces to \be \psi\sim \frac{1}{r} \exp
\left(-r/a_s\right),\label{5}\ee where $a_s=
(m\varepsilon_b)^{1/2}$ is the scattering length. This
parametrization is very convenient for the system of cold
fermionic atoms where the scattering length is a tunable parameter
\cite{27}. However, for quarks  the scattering length is an
ill-defined  quantity.

Let us see, then, what happens to the wave function (\ref{4}) in
ultrarelativistic case if $m\to  0$ in line with chiral symmetry
restoration. Then \be \psi\sim \frac{1}{r} \exp \left(-
\frac{r}{2} (-\varepsilon_b^2)^{1/2}\right ).\label{6}\ee This
result shows that the system develops an instability due to the
coalescence of the quark and antiquark branches \cite{21}.
The arguments presented here are based  on the asymptotic form (\ref{4})
of the wave function. To investigate the vicinity of the instability
point one should consider    the system of coupled equations
for positive and negative  energy components
written down in \cite{21}. This will be done in a forthcoming paper.
Therefore the  chiral symmetry restoration in $(NM)\to (QM)$
transition is a subtle point which deserves further investigation.

Next we turn from the crossover parameter to the Ginzburg-Levanyuk
number $Gi$ \cite{28}. The two quantities are interrelated since
both characterize the fluctuating pairs. Let us start with the
expression for $Gi$ in case of a clean three-dimensional
superconductor \cite{28} \be Gi=\frac{27\pi^4}{28\zeta(3)}
\left(\frac{T_c}{E_F}\right)^4\simeq
80\left(\frac{T_c}{E_F}\right)^4,\label{7}\ee where
$\zeta(3)\simeq 1.2,~~ T_c$ is the critical temperature and $E_F$
-- the Fermi energy. We note in passing that in \cite{29} $Gi$ was
underestimated by two orders of  magnitude. This means an extreme
narrowness of the fluctuation region. If we apply (\ref{7}) to the
quark system, take $\mu$ for $E_F$ and put $\mu\simeq (0.3-0.4) $
GeV, $T_c\simeq(0.04-0.05)$ GeV \cite{3,6,7}, we obtain \be Gi
\gtrsim 10^{-2}\label{8}\ee which is a huge number as compared to
that for ordinary superconductors. It is by two orders of
magnitude larger than the value of $Gi$ for quarks presented in
\cite{29}.

At this point one may ask to what extent is the equation (\ref{7})
for $Gi$ applicable to  the quark system. One of the ways to
derive it \cite{30} is to compare the radius of interaction
between fluctuations with the correlation radius in the system far
from the transition point. Then to arrive at (\ref{7}) use is made
of the Ginzburg-Landau functional. Therefore the estimate
(\ref{9}) relies on the applicability of the Ginzburg-Landau
theory to the quark system at moderate density. We shall discuss
this point at the end of the paper.

Next we wish to express $Gi$ in terms of the crossover parameter.
We remind the expression for the coherence length \cite{28}
\be
\xi^2=\frac{7\zeta(3)}{48 \pi^2 T^2_c} v^2_F,
\label{9}
\ee
where $v_F$ is the velocity at the Fermi surface.
 Using Eqs. (\ref{7}) and (\ref{9}) we obtain
 \be
 Gi=\frac{21\zeta(3)}{64} (k_F\xi)^{-4} \simeq \frac{5\cdot
 10^{-2}}{(n^{1/3}\xi)^4}.\label{10}\ee
 We see that the fluctuation effects  in  a quark system at
 moderate density are very strong. Next we have  to determine what
 is  the  dominant type of fluctuations -- that of the order
 parameter, or of the gauge field potential. The fluctuation
 contribution to the free energy can be estimated \cite{28}
 dividing the volume of a speciman by the cube of the fluctuation
 correlation length of the given type of fluctuation.
 The correlation length of the order parameter is, as we have
 seen, $\sim 1$~fm, while that of the gluon field is $\sim 0.2$
 fm\footnote{Strictly speaking, within the local Landau-Ginzburg
 theory we may consider only fluctuations with the correlation
 length larger than the pair size.} \cite{31}. Therefore
 fluctuation of the gauge field dominate.

\section{The Ginzburg-Landau Functional}

The standard framework to consider fluctuations  is the
Ginzburg-Landau functional (GLF). In this section we shall present
the derivation of the GLF for quarks starting from the general
form of the effective action common for the NJL, gluon exchange,
or instanton models. Considerations presented below may be
regarded as an extension of our paper \cite{11}.

We start with the QCD Euclidean partition function \be Z=\int
DAD\bar \psi D\psi \exp (-S),\label{11}\ee where
 \be
  S=\int d^4x\bar \psi(-i\gamma_\mu D_\mu-im +i\mu\gamma_4)\psi +
  \frac14 \int d^4 x F^a_{\mu\nu} F^a_{\mu\nu}.\label{12}\ee

  In (\ref{12}) color and flavor indices are suppressed, $N_f=2,
  N_c=3$, and the chemical potential $\mu$ is introduced.
  Performing integration over the gauge fields one gets effective
  fermion action in terms of cluster expansion \cite{13, 31}
  \be
   Z=\int D\bar \psi D\psi \exp (-\int d^4 x L_0
   -S_{eff}),\label{13}\ee
   with $L_0 =\bar \psi (-i\gamma_\mu \partial_\mu -im
   +i\mu\gamma_4)\psi$ and effective action \\
   $S_{eff}
   =\sum^\infty_{n=2} \frac{1}{n!} \ll\theta^n\gg$, where $\theta
   =\int d^4 x\bar \psi (x) g\gamma_\mu A^a_\mu (x) t^a \psi (x)$
   and the double brackets denote irreducible cumulants \cite{32}.

   The derivation of the GLF from the effective action entering
   into (\ref{13}) is a complicated task for further work. Only the
   first step in this direction has been done recently \cite{13}.
   Here we shall follow much simpler route which is used to derive
   the GLF for electronic superconductors \cite{33}.

   We replace $S_{eff}$ in (\ref{13}) by an effective four-fermion
   interaction with symmetry properties of the two-flavor QCD.
   This might be either NJL, or instanton vertex, or contact
   interaction imitating one-gluon exchange. Symbolically such
   interaction looks like
   \be
   L_{int} =g (\bar \psi\hat K\psi) (\bar \psi \hat K
   \psi),\label{14}\ee
   where the constant $g$ has a dimension $m^{-2}$.

   Then we perform the Fierz transformation in the Lorentz, color
   and flavor spaces. As a result the  interaction term takes the
   form
   \be
   L_{int} =g (\bar \psi^c \hat R \psi) (\bar \psi \hat R^+
   \psi^c).\label{15}\ee
   Here $(\bar \psi^c\hat R\psi)$ is a scalar  diquark  in color
   $\bar 3$ state
   \be
    \bar \psi^c\hat R \psi =\psi^T_{\alpha i} C\delta_{\gamma3}
    \varepsilon_{\alpha\beta\gamma} (\tau_2)_{ij} \gamma_5
    \psi_{\beta j},\label{16}\ee
    where $C=\gamma_2 \gamma_4$ and the presence of the color
    structure $\delta_{\gamma3}\varepsilon_{\alpha\beta\gamma}$
    signals that color symmetry is broken by such diquarks.

    Next step is to integrate the partition function over quark
    fields. The interaction is quartic in the fermion fields and
    therefore one has to apply the Hubbard-Strotonovich  trick
    (bosonization)
   $$
     \exp \left\{ g\int^\beta_0 d\tau \int d\ver (\bar
     \psi^c\hat R \psi) (\bar \psi \hat R^+ \psi^c)\right\}=
$$
\be
 =\int D\Delta^* D\Delta \exp\left\{-\int^\beta_0 d\tau \int d\ver\left[ \frac{|
 \Delta|^2}{g} - \Delta  (\bar \psi \hat R^+ \psi^c)- \Delta^*(\bar
     \psi^c\hat R \psi)\right] \right\},\label{17}\ee
     where $\Delta$ is a complex scalar field. From the Lagrange
     equation of motion for $\Delta^*$ we see that $\Delta= g\lan
     \bar \psi^c \hat R \psi\ran$, so that $\Delta$ has a
     dimension of mass as it should be for the gap parameter. Now
     we can integrate the partition function over the quark fields
     and obtain
     \be
Z=\int D\Delta^* D \Delta \exp
\left\{-\frac{1}{g} \int^\beta_0 d\tau \int d\ver
 |\Delta|^2 - \Omega_B(T,\mu, \Delta^*, \Delta) \right\},\label{18}\ee
 where $\Omega_B$ is the Bogolubov functional
 \be
  \Omega_B =-\frac12 tr \ln \left( \begin{array}{cc}
  \Delta \hat R&i\partial_\mu\gamma_\mu -i\mu\gamma_4\\
  -i\partial_\mu^T\gamma_\mu^T+i\mu\gamma_4^T& \Delta^*
   \hat R^+\end{array}\right).\label{19}\ee

We use the following representation
\be
\vegamma=\left( \begin{array}{ll}
0&-i\vesig\\
i\vesig& 0\end{array} \right),~~\gamma_4 = \left(
\begin{array}{ll} 1&0\\ 0&-1\end{array} \right),\label{20}\ee
\be
C=\gamma_2\gamma_4,~~ C^{-1}\gamma_\mu C =-\gamma\mu^T,~~
\partial^T_\mu=\overleftarrow{\partial}_\mu=-
\overrightarrow{\partial}_\mu.\label{21}\ee
 In (\ref{19}) the quark mass is for simplicity set to zero. The
 important assertion  we are making is that the gauge field may be
 introduced  at the  last step by replacing the gradient by the
 covariant  derivative and adding the  Yang-Mills part of the
 Lagrangian.
   We may also
  recall \cite{6,7,11} that the Fierz transformation from
  (\ref{14}) to (\ref{15}) results in some extra terms in
  (\ref{15}) including the chiral condensate  ($\bar \psi_{\alpha i}
  \delta_{\alpha\beta} \delta_{ij} \psi_{\beta j})$.
 It has been shown in \cite{11} that in the leading
  approximation the thermodynamics of the system is determined by
  diquarks.  Finally, we  wish to note that according  to (\ref{18})
  the partition function  is obtained by averaging the  Bogolubov
  functional
  $\Omega_B$ with the Gaussian weight factor  $|\Delta|^2/g$.

  Next we rearrange (\ref{19}) using the identity
\be tr \ln \left( \begin{array}{ll}
 B&A\\-A^T&B^+\end{array}\right) =tr \ln AA^T +tr \ln \{ 1+A^{-1}
 B(A^T)^{-1} B^+\}.
 \label{22}\ee

 Before we apply (\ref{22}) to $\Omega_B$ given by (\ref{19}), we
 rewrite the operator $\hat R$ (see (\ref{16})) as $\hat R=\hat
 R_0C$,  where $C=\gamma_2\gamma_4$ is the charge conjugation
 operator. Then the following formula is easily worked out to be
 \be
 A^{-1} \Delta \hat R (A^T)^{-1} \Delta^* \hat R^+ = \hat R_0 \hat
 R_0^+ \sum_p \Delta (p) \Delta^* (p) \sum_k G(k)
 G(p-k),\label{23}\ee
 where $A=i\partial_\mu \gamma_\mu -i\mu \gamma_4,$ and $G$ is the
 thermal propagator
 \be
 G(q) =\frac{i}{\veq \vegamma + q_4 \gamma_4 + i\mu \gamma_4}
 \equiv \frac{i}{\hat q+ i\mu\gamma_4}\label{24}\ee
 with $q_4=-\pi (2n+1) T, T=\beta^{-1}$. Next we return to formula
 (\ref{18}) and write
 \be ZZ_0^{-1} =\exp \left\{ - (\Omega - \Omega_0)\right\},
 \label{25} \ee
with $(\Omega - \Omega_0)$ being the $\Delta\neq 0$ part of the
thermodynamic potential. Now we can finally combine
(\ref{18})-(\ref{19}) and (\ref{23})-(\ref{25}) and somewhat
symbolically write \be \Omega-\Omega_0 =\frac{1}{g} \int^\beta_0
d\tau \int  d\ver |\Delta|^2 -\frac{1}{2} tr \ln \left\{ 1+ \hat
R_0 \hat R_0^+ \Delta \Delta^* GG \right\}.\label{26}\ee
Expanding the logarithm in (\ref{26}) in powers of $\Delta^2$ we
arrive at the GLF. Following the standard approach we take into
account the spatial variation of $\Delta$ in the quadratic term
and neglect such dependence in the quartic term. The diagrams
corresponding to the terms proportional to  $|\Delta|^2$ and
$|\Delta|^4$ are shown in Fig.1.

\begin{figure}
\centerline{\includegraphics[width=12cm,
keepaspectratio=true]{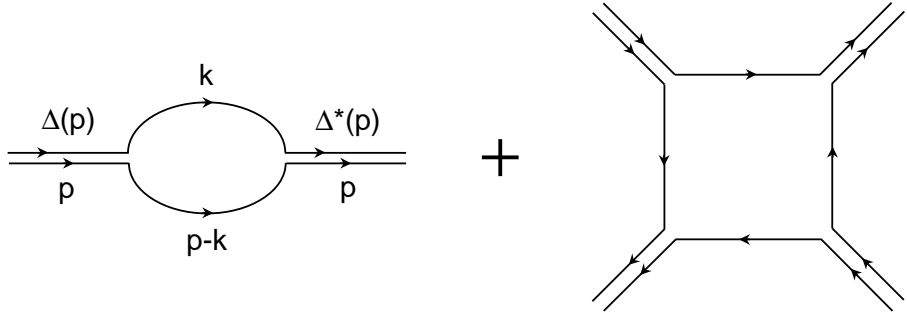}} \caption{Quadratic and quartic
terms of GLF.} \label{fig1}
\end{figure}

We start with the quadratic term in (\ref{26}). It reads \be
\Omega^{(2)} =\frac{1}{g} \int^\beta_0 d\tau \int d\ver |\Delta|^2
-\frac12 tr \hat R_0 \hat R_0^+ \sum_p \sum_k \frac{-\Delta(p)
\Delta^*(p)}{(\hat k+ i\mu\gamma_4) (\hat p -\hat k + i\mu
\gamma_4)}.\label{27}\ee

The trace over flavor and color indices is \be tr_{f,c}  \hat R_0
\hat R_0^+ =N_f (N_c-1) =4.\label{28}\ee
After this trace is taken, formula (\ref{27}) still contains $tr'$
over the Lorentz indices ($\gamma$-matrices).

The GLF corresponds to the 'soft loop" approximation \be tr'\sum_k
\frac{-1}{(\hat k+i\mu\gamma_4)(\hat p-\hat k+i\mu\gamma_4)}
\simeq A+B\vep^2.\label{29} \ee

First we compute the static term $A$. We have \be
A=\frac{1}{\beta} tr' \sum_{\omega_n} \int \frac{d\vek}{(2\pi)^3}
\frac{1}{(\hat k+i\mu\gamma_4)(\hat k-i\mu\gamma_4)}.\label{30}\ee
By simple algebra it is easily seen that \be tr' \frac{1}{(\hat
k+i\mu\gamma_4)(\hat k-i\mu\gamma_4)}=2\left\{
\frac{1}{k^2_4+(k-\mu)^2}+\frac{1}{k^2_4+(k+\mu)^2}\right\}.
\label{31}\ee
 We assume that one can neglect the contribution of antiparticles.
 This means that only the first term should be kept in the
 representation (\ref{31}). Then we perform  the Matsubara
 summation in (\ref{30}) and obtain
 \be
 A=\frac{\nu_F}{4} \int^\Lambda_{-\Lambda} \frac{d\xi}{\xi} th
 \frac{\xi}{2T}.\label{32}\ee
 Here $\nu_F=2\mu^2/\pi^2$ is the density of states at the Fermi
 surface. Due to color and flavor degrees of freedom it is four
 times larger than the corresponding BCS theory factor. Parameter
 $\Lambda$ is the cut-off  of the four fermion interaction. In BCS
  theory it is the Debye frequency, in the NJL model
  $\Lambda\simeq 0.8 $ GeV.  As we shall see, the value of
  $\Lambda$ is irrelevant as soon as $\Lambda/2T_c\gg1.$
  Integration in (\ref{32}) is over  $\xi=k-\mu$.

  To proceed further we make an  important assumption. In the
  summation over $p$ in (\ref{27}) we integrate over Fourier
  coefficients with $\omega\neq  0.$ In BCS theory this leads to
  small corrections to the
GLF coefficients \cite{33}. This means that we keep in (\ref{27})
only $ \Delta(\vep)$ and $ \Delta^*(\vep) $ having $\omega =0$
frequency.
Then \be \sum_p \Delta (p) \Delta^* (p) =\beta\int
\frac{d\vep}{(2\pi)^3} \Delta (\vep) \Delta^* (\vep) =\beta \int
d\ver |\Delta (\ver)|^2. \label{33} \ee
Under the same assumption we can integrate over $\tau$ the first
term in (\ref{27}). Next we expand the integrand in (\ref{32}) in
powers of $(T_c-T)/T_c$ \be \frac{1}{\xi} th \frac{\xi}{2T}
\simeq\frac{1}{\xi} th \frac{\xi}{2T_c} + \frac{T_c-T}{2T_c^2}
\left( ch\frac{\xi}{2T_c}\right)^{-2}.
 \label{34}
  \ee

  Upon insertion into (\ref{27}) and with the account of
  (\ref{28}) the first term in the right-hand  side of (\ref{34})
  cancels with the first term in the right-hand side of
  (\ref{27}). This is because of the gap equation or equivalently
  from the definition of the  critical temperature. Then we
  perform the integration in (\ref{32}) keeping only the last term
  in (\ref{34}) and setting the limits  of integration to $\mp
  \infty$ since $\Lambda/2T_c\sim 10$ and the  integrand is a
  rapidly  decreasing function. The result for $A$ reads
  \be
  A=\frac{\nu_F}{2} \left( \frac{T_c-T}{T_c}\right).
  \label{35}
  \ee

   Now we turn to the coefficient $B$ in (\ref{29}). Keeping in
   mind the assumption formulated prior to Eq. (\ref{33}), we
   write\footnote{In fact the above assumption was tacitly used
   already in Eq. (\ref{29}).}

   $$ B=\frac{1}{2\vep^2\beta} tr' \sum_{\omega_n} \int
   \frac{d\vek}{(2\pi)^3} \frac{1}{(\hat k + i\mu\gamma_4)}
   \left(\vep \frac{\partial}{\partial\vek}\right)^2
   \frac{1}{(\hat k-i\mu\gamma_4)}=$$
   \be
  = -\frac{\nu_F}{48\beta} tr'\vegamma^2 \sum_{\omega_n}
   \int^{+\infty}_{-\infty} \frac{d\xi}{(\xi^2+k^2_4)^2} =- \nu_F
   \frac{7 \zeta(3)}{96 \pi^2T^2_c}.\label{36}\ee
With the coefficients $A$ and $B$ at hand we can write down the
quadratic part of the thermodynamic potential \be \Omega^{(2)}
=\beta\nu_F \int d\ver \left\{ t|\Delta ( \ver)|^2 +
\frac{7\zeta(3)}{48 \pi^2T^2_c} | \grad
\Delta(\ver)|^2\right\},\label{37}\ee where $t=(T-T_c)/T_c$.

The derivation  of the quartic term essentially repeats that of the
quadratic term presented above. Therefore we omit the technical
details. The result reads
\be \Omega^{(4)} =\beta\nu_F
\frac{7\zeta(3)}{16 \pi^2T^2_c} \int d\ver |
\Delta(\ver)|^4.\label{38}\ee
Now we can assemble the pieces
together. We remind the two assumptions formulated above: (i)
non-zero Matsubara are decoupled, and (ii) the gluon field enters
via the covariant derivative. The result reads
\be \Omega=\beta
\int d\ver F(T, \mu, \Delta^*, \Delta, \mathbf{A}^l),\label{39}\ee
\be F=\nu_F \left\{ t|\Delta|^2 +\frac{\beta}{2} |\Delta|^4
+\gamma |\mathbf{D} \Delta|^2\right\} - \frac{1}{2} A_k^l \vec{\mathbf{\nabla}}^2
A_k^l.\label{40}\ee
Here
\be
\mathbf{D} =\mathbf{\nabla} -ig \frac{\lambda^l}{2}
\mathbf{A}^l,\label{41}\ee
\be \beta =\frac{7\zeta (3)}{8
\pi^2T^2_C},~~ \gamma =\frac{7\zeta (3)}{48
\pi^2T^2_C}.\label{42}\ee
Note that $\gamma=\xi^2$, where $\xi^2$
is the BCS theory correlation length (see Eq. (8)).
The last term in (\ref{40}) is the contribution of the gluon field in the
Coulomb gauge with cubic and quartic terms neglected.

One may argue that expression (\ref{40}) contains nothing new. It
coincides with the standard   GLF with the coefficients
(\ref{42}) obtained by Gorkov almost half a century ago \cite{34}.
For the quark system the GLF was presented in \cite{1,35} and some
later papers. What we have tried to do here is to expose several
subtle points in the derivation of the GLF for quarks. It should
be clear that even if one takes for granted that in some
mysterious way the role of the gluon sector reduces to the
generation of an effective four-quark interaction it is by far not
enough to straightforwardly derive the GLF.

\section{Fluctuation Induced Color Diamagnetism}

We have seen that the $(NM)\to (QM)$ transition takes place in the
strong fluctuation regime. Fluctuating diquarks are the precursors
for the Cooper pairs forming the BCS condensate. Below we consider
fluctuations of the gluon field. This type of fluctuations is
responsible for two effects \cite{36}: (i) the lowering of the
critical temperature (fluctuation diamagnetism), and (ii)
modification of the order of the phase transition (first order
instead of second). Both points with regard to quark system have
been discussed in recent years \cite{37,38}.

Our emphasis here will be on another aspect of the gluon field
fluctuation phenomena, namely "the emerging phenomenology of $\lan
(A^a)^2\ran$" gluon condensate \cite{39,40}. In the last few years
there has been a growing interest in condensates of dimension two
\cite{41}. It might be that there is some (indirect) connection
between the phenomenology of $\lan A^2\ran$ and gluon field
fluctuation in dense quark matter. There is a further possibility
that the problem of $\lan A^2\ran $ is linked to the Anderson
transition in quark matter. In order to clearly define what we
mean by $\lan A^2\ran $ in the context of dense quark matter we
start with the derivation of the basic equations.

The partition function expressed in terms of the variables
$\Delta, \Delta^*$ and $\mathbf{A}^l$ reads
$$Z=\int D\Delta^* D\Delta D\mathbf{A}^l
\exp \left\{ -\beta \int d\ver F (T,\mu, \Delta^*, \Delta,
\mathbf{A}^l)\right\}=$$ \be = \int D\Delta^* D\Delta
Z_A,\label{43}\ee
where $F$ is the GLF (\ref{40}) and
\be Z_A
=\int D\mathbf{ A}^l \exp \left\{ -\beta \int d\ver
F\right\}.\label{44} \ee
Integration over the gauge fields yields $$ Z_A = \exp \{ -\beta
\nu_F \int d\ver (t |\Delta|^2 +\frac{\beta}{2} |\Delta|^4)+
$$
\be +\frac{1}{\beta} \int \frac{d\vek}{(2\pi)^3} \ln (2\tilde
\gamma g^2 |\Delta|^2 + \vek^2)\}.\label{45}\ee
Here $\tilde \gamma = \nu_F \gamma$. The coupling  constant $g^2$
in (\ref{45}) includes the averaging of the kinetic term
$|\mathbf{D} \Delta |^2$ over the genetators $\lambda^l$. The
corresponding calculations may be found in Ref. \cite{42}. Taking
the derivative of $\ln Z_A$ with respect to  $|\Delta|$ we obtain
$$
-\frac{T}{2V} \frac{\partial}{\partial|\Delta|}\ln Z_A =\nu_F \{
t|\Delta| +\beta |\Delta|^3\}+$$
\be + T\int
\frac{d\vek}{(2\pi)^3} \frac{2\tilde \gamma g^2|\Delta|}{\vek^2 + 2
\tilde \gamma g^2 |\Delta |^2}.\label{46}\ee

The same derivative calculated directly from (\ref{44}) and
(\ref{40}) reads
$$-\frac{T}{2V} \frac{\partial}{\partial|\Delta|}\ln Z_A =\nu_F \{
t|\Delta| +\beta |\Delta|^3\}+$$
$$ + \gamma g^2 |\Delta| Z_A^{-1} \int D \veA^l (\veA^2)
\exp (-\beta \int d \ver F)\} = $$
\be = \nu_F \{ t |\Delta| + \beta |\Delta|^3 + \gamma g^2 |\Delta|
\langle\veA^2 \rangle \}. \label{47}\ee
Comparing (\ref{46}) and (\ref{47}), one obtains
$$\langle \veA^2 \rangle = Z^{-1}_A \int D \veA^l (\veA^2) \exp
\{-\beta \int d\ver F\} =$$
\be = 2T \int \frac{d\vek}{(2\pi)^3} \frac{1}{\vek^2 + M^2},
\label{48} \ee
where $M^2 = 2 \nu_F \gamma g^2 |\Delta|^2$. We recognize in $M$
the London penetration depth. The average $\langle \veA^2 \rangle$
is the expectation value of $\veA^2$ in a fixed bosonic field
$\Delta$. Next we integrate (\ref{47}) back over $\Delta$ and
obtain
\be F = \nu_F \left\{ \left( t + g^2 \xi^2 \langle \veA^2
\rangle \right) |\Delta|^2 + \frac{\beta}{2} |\Delta|^4
\right\}.\label{49} \ee
The critical temperature shifted due to gluon field fluctuations
corresponds to zero of the quadratic term coefficient \cite{36,
28, 43}
\be T'_C = T_C(1 - g^2 \xi^2 \langle \veA^2 \rangle). \label{50}
\ee
We can estimate $g^2 \xi^2$ as $g^2 = 4 \pi \alpha_S \simeq 4$,
$\xi \simeq 2$~fm, $g^2 \xi^2 \simeq 400$~GeV$^{-2}$. Then the
formal absolute upper bound on $\langle \veA^2 \rangle$ following
from (\ref{50}) would be $\langle \veA^2 \rangle < (50$ MeV)$^2$.
Later we shall return to the estimate of $\langle \veA^2 \rangle$.

The second effect of the gauge field fluctuations is the
replacement of the second order phase transition by the first
order one. This point was thoroughly discussed in the literature
\cite{36, 1, 37, 38,11}. A careful look at the expression (\ref{49})
for $\langle \veA^2 \rangle$ shows that it induces the
$|\Delta|^3$ term in the GLF. To see this consider equation
(\ref{48}) for $\langle \veA^2 \rangle$ and expand it in terms of
$M/\Lambda$, where $\Lambda \simeq 0.8$~GeV\footnote{According to \cite{37}
$\Lambda = T_C$ (in the high density regime).} is the cut-off
parameter.
\be \langle \veA^2 \rangle = \frac{T \Lambda}{\pi^2} \left\{ 1 -
\frac{\pi}{2} \frac{M}{\Lambda} + \frac{M^2}{\lambda^2} - \ldots
\right\} \simeq \frac{T \Lambda}{\pi^2} - \frac{g}{\pi^2} T
(\mu \xi) |\Delta|.\label{51} \ee
Then instead of (\ref{49}) we have
\be F = \nu_F \left\{ \left( t + g^2 \xi^2 \frac{T \Lambda}{\pi^2}
\right) |\Delta|^2 - \frac{g^3}{\pi^2} T \mu \xi^3 |\Delta|^3 +
\frac{\beta}{2} |\Delta|^4 \right\}.\label{52} \ee
Due to the term proportional to $|\Delta|^3$ the potential $F$
acquires a second minimum at a finite value of $|\Delta|$. This
means that the phase transition is of the first order. For further
discussion of this question see \cite{37,38}.

From (\ref{51}) we obtain another upper bound on $\langle \veA^2
\rangle$
\be \langle \veA^2\rangle < \frac{T_C \Lambda}{\pi^2} \simeq (60
\mbox{MeV})^2 \label{53} \ee
for $T_C = 50$~MeV, $\Lambda = 800$~MeV.

We see that the $(NM) \rightarrow (QM)$ transition is accompanied
by the gluon field fluctuations with the typical magnitude
$\langle \veA^2\rangle \lesssim (50 \mbox{MeV})^2$. What might be
the meaning of this result? Is this number large or small? The
quantity $\langle \veA^2\rangle$ is gauge-variant, but in
\cite{40} it was shown that it attains its minimum in the Coulomb
gauge. The comparison of the above result with lattice
calculations of $\langle A^2_{\mu} \rangle$ at zero baryon density
below and above $T_C$ \cite{39} shows that our value is by more
than an order of magnitude smaller than the lattice result. One
may argue that the two quantities are defined somewhat differently
(see \cite{39,40}). The physical reason for possible mismatch is
that the increase of the baryon/quark density leads to the
suppression of the gluon field \cite{10,13}. However the
suppression by a factor of 10 - 20 seems to contradict the
expected magnitude of the effect \cite{10}. The possible way to
resolve this difficulty (if it exists) is to return to the general
expression for the GLF (\ref{40}) and reanalyze the loop diagram
which gives rise to the coefficient $\gamma$ in (\ref{40}) and
then to the coefficient $\xi^2$ in front of $\langle
\veA^2\rangle$ in (\ref{50}).

\section{A Marginal Remark on Anderson \\
Localization}

Anderson localization $(AL)$ \cite{44} is one of the basic
concepts of contemporary condensed matter physics. Recently it was
realized \cite{45,28} that this phenomena is important for the
description of superconductivity in strongly disordered systems.

The explanation of $AL$ can be given in terms of the two-particle
Green's function (\ref{29}) (the loop diagram in Fig.1). This
diagram embedded into the disordered background determines the
dynamical momentum dependent diffusion coefficient which turns
zero at the localization edge. In the theory of superconductivity
the disorder is created by impurities. According to Ioffe-Regel
criterion \cite{46} the phase transition to the $AL$ regime takes
place at \be k_Fl \lesssim 1, \label{54} \ee where $l$ is the mean
free path. Approaching the localization region the coefficient
$\gamma$ in the $GLF$ (\ref{40}) (and hence the coefficient
$\xi^2$ in (\ref{50})) undergoes a drastic change \cite{45}. In
particular, close to mobility edge (condition (\ref{54})) the
coefficient $\gamma = \xi^2$ in (\ref{40}) is substituted by
\cite{45} \be \gamma = \left( \frac{D_0 l}{T_C} \right)^{2/3},
\label{55} \ee where $D_0 = \frac{1}{3} v_F l$ is the Drude
diffusion coefficient. Formula (\ref{55}) is one of several
possible expressions for $\gamma$. The concrete scenario of the
evolution of $\gamma$ as a function of the disorder strength
depends upon the values of the critical exponent and parameters
(\ref{54}) -- see Refs. \cite{28,45}. Using (\ref{55}) we may
illustrate the idea of possible presence of $AL$ in the quark
system in the transition region and estimate the renormalization
of the coefficient $\gamma$ in $GLF$.

The hypothesis that $AL$ may play some role in quark systems was
first formulated in \cite{47}. The role of impurities was
attributed to random components of the gluon fields. The behavior
of the quark Green's function was analyzed and it was shown that
localization should not be confused with confinement. In the
framework of this idea let us estimate the renormalization of the
$GLF$ coefficient $\gamma$ according to (\ref{55}). The
characteristic scale of the quark mean free path is $l \simeq
1$~fm. This might be, e.g., the average distance between
instantons in the instanton vacuum picture of $QCD$
\cite{48}\footnote{Because of quark spin-flip instantons may be
similar to magnetic impurities.}. For $l = 1$~fm, $v_F = 1$,
$T_C=50$~MeV equation (\ref{55}) yields $\gamma \simeq
1.2$~fm$^2$, while $\gamma = \xi^2$ with $\xi = 2$~fm gives
$\gamma \simeq 4$~fm$^2$. Approaching the edge of localization the
expression (\ref{51}) and the estimate (\ref{53}) are not valid
any more since the inverse penetration depth $M$ in (\ref{51})
becomes energy dependent through the energy dependence of $\gamma$
(expression (\ref{55}) is valid only in the immediate vicinity of
the mobility edge). With the above estimate for $\gamma$ we may
return to Eq. (\ref{50}) and obtain a new upper bound on $\langle
\veA^2 \rangle$ which is $\langle \veA^2 \rangle \lesssim
(100$~MeV$)^2$. Hence the discrepancy between our estimate for
$\langle \veA^2 \rangle$ and that given in Ref.\cite{39} may be at
least partly eliminated by the account of $AL$.

\section{Summary}

It is clear that further work is needed before the process of
quark matter formation is finally understood. Three important
features of this process have been examined in the present paper.
They are: (i) crossover from strong coupling/low density
to weak coupling/high density,
(ii) strong fluctuation regime, (iii) possible Anderson
localization. We have seen that these three points are
interrelated.

It can be definitely stated that there is no direct transition
from nuclear matter phase to quark superconducting $BCS$-like
phase. In between the two phases there is a crossover region with
strong fluctuations and possible Anderson localization. Despite
the extensive investigations our understanding of the $(NM) \rightarrow
(QM)$ transition is in no way being complete. In particular, the
behavior of the string tension in this region is almost
intractable (see, however, Ref.\cite{13}).

We have presented a detailed derivation of the $GLF$ within the
mean-field three-dimensional effective theory. We have found that
the $GL$ approach is on a shaky ground in this region. Small
correlation length of the gluon field makes the local
approximation baseless, fluctuation corrections are large and the
influence of antiquarks is not negligible. Therefore the
investigation of the interior of neutron stars using $GLF$ (see,
e.g., \cite{49}) can hardly bring reliable results.

We have found an interesting intersection between the $QM$ physics
and Anderson localization in highly disordered media. The behavior
of the quark loop diagram shown in Fig.1 embedded into the gauge
field background should be studied more deeply.

Finally we wish to mention that the location of the crossover
region on the $QCD$ phase diagram has been discussed in a recent
publication \cite{50}.

\section{Acknowledgments}

We would like to thank N.O.Agasian, F.V.Gubarev, V.I.Shevchenko
and  Yu.A.Simonov for discussions and critical remarks. This work
was partially supported by the RFBR grant 06-02-17012, Leading
Scientific Schools grant \# 843.2006.2 and State Contract
02.445.11.7424. 2006-112. Finally, we wish to thank Prof. H.Abuki
for bringing our attention to Ref [51], which contains a
comprehensive analysis of several questions discussed in the
present paper. We are also grateful to Prof. N.Itoh, whose paper
[52] was not known to us before.

\end{document}